\documentclass[useAMS,usenatbib,usegraphicx]{mn2e}
\usepackage{times}

\title[Milky Way dust extinction measured with QSOs ]{Milky Way dust extinction measured with QSOs }

\author[C. Wolf]{Christian Wolf
\smallskip \smallskip \\
Research School of Astronomy and Astrophysics, Australian National University, Canberra, ACT 2611, Australia, email: christian.wolf@anu.edu.au \\
}

\begin{document}
\date{accepted 30 Sep 2014}
\maketitle

\begin{abstract}
We investigate reddening by Milky Way dust in the low-extinction regime of $E_{B-V}<0.15$. Using over 50,000 QSOs at $0.5<z<2.5$ from the SDSS DR7 QSO Catalogue we probe the residual SDSS colours after dereddening and correcting for the known spectroscopic redshifts. We find that the extinction vector of Schlafly \& Finkbeiner (2011, SF11) is a better fit to the data than that used by Schlegel et al. (1998, SFD). There is evidence for a non-linearity in the SFD reddening map, which is similarly present in the V1.2 map of the {\it Planck} Collaboration. This non-linearity is similarly seen when galaxies or stars are used as probes of the SFD map. 
\end{abstract}


\section{Introduction}

Interstellar dust is ubiquitous and literally clouds our view of the Universe beyond the most nearby stars, although visual absorption is below 10\% on a third of the sky. Its main impact on distant galaxies and quasars is a bandpass-dependent absorption that changes colours and luminosities to a degree that varies across the sky. The interstellar dust associated with our Milky Way is sufficiently nearby so that it alters the appearance of all extragalactic objects, e.g. galaxies, quasars and supernovae therein, independently of their distance. However, stars within the Milky Way are affected only by the dust in front of them, so their appearance depends on their location and the 3D-cloud structure of the dust \citep[e.g.][]{G14}. Extragalactic objects are also affected by dust in the outskirts of galaxies that are close to the line-of-sight \citep{Menard10}, although for most objects this effect is negligible in comparison to the Milky Way's stronger foreground extinction.

Accurate knowledge of dust extinction is crucial for a whole range of extragalactic studies that rely on measurements of accurate spectral energy distributions (SEDs) and luminosities. It is thus also crucial for the accurate measurement of distances where SEDs and luminosities are used to infer them. Indeed standard candles, e.g. type-Ia supernovae, are used to derive redshift-distance relations, map the expansion history of the Universe and constrain cosmological parameters. 

It has now been asserted that uncertainties in the Milky Way dust extinction are the second-largest source of systematic uncertainties in the measurement of cosmological parameters from type-Ia supernova samples \citep{Bet14}. The primary source is the photometric calibration of supernova observations including an imperfect knowledge of filter transmission curves used in the analysis. As these primary uncertainties will be much reduced in the next generation of supernova surveys, e.g. the SkyMapper Supernova Survey \citep{SMSNS}, our understanding of Milky Way dust extinction is the next-most important factor in the line. For this reason alone, and for several others, we would like to improve our understanding of dust extinction.

Most currently cited extragalactic work is based on the dust maps and extinction vector by \citet{SFD98}, hereafter SFD98. This map of estimated dust opacity was calibrated such that the reddening of nearby elliptical galaxies, as expressed in $E_{B-V}$, is on average correctly described. However, more recently the accuracy of the SFD98 dust map and reddening vector have been questioned. The works of \citet{S10}, hereafter S10, and \citet{SF11}, hereafter SF11, have proposed a new reddening vector, which in conjunction with the SFD98 dust map makes the de-reddened SDSS colours of Milky Way stars more consistent. And using colours of SDSS galaxies, \citet{Ya07} found unexpected non-linear trends with reddening, which might be explained if the galaxies themselves biased the SFD98 map by adding to the dust emission from which the extinction is derived. Also, \citet{Peek13} has investigated a non-linearity in the SFD98 dust map that manifests as a trend in the de-reddened $g-r$ colour residuals of elliptical galaxies. The feature seemed to disappear when using the first reddening map (V1.1) by the {\it Planck} team, who have since published an update \citep[V1.2,][]{Planck}.

In this paper, we revisit these issues using SDSS quasars (QSOs) and their de-reddened colour residuals. Sect.~2 details our sample and methods for de-reddening the measurements. Sect.~3 investigates the colour residuals when applying two currently considered extinction vectors, those of SFD98 and SF11, and several alternative $E_{B-V}$-maps. Sect.~4 discusses a North-South asymmetry seen in SDSS stars from the point of view of QSOs, and Sect.~5 discusses apparent non-linearities in the reddening map.

\section{Data and de-reddening}

\subsection{The SDSS QSO sample}

We use the data release 7 (DR7) of the SDSS QSO catalogue by \citet{DR7QSO}, which contains over 105,783 spectroscopically identified QSOs, mostly in the Northern Galactic Cap. The area of the Northern Galactic Cap has been consistently calibrated using an {\it ubercal} approach \citep{ubercal}, while the supplemental areas at Southern Galactic latitudes cannot be put onto the same calibration given it has no overlap with the main area. Indeed, \citet{S10} find a North-South asymmetry in the colours of blue-tip stars and discuss SDSS calibration differences as one possible origin. We thus restrict the sample to the area of the Northern Galactic Cap for our main analysis, but compare the Northern and Southern samples in Sect.~4.

\begin{figure}
\centering
\includegraphics[clip,angle=270,width=0.985\hsize]{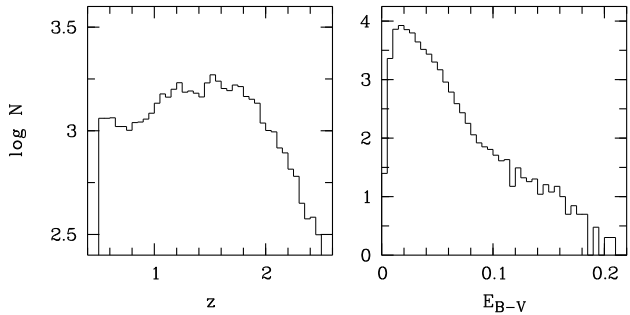}
\caption{{\it Left:} Redshift histogram of Northern SDSS QSO sample used in this paper.
{\it Right:} Reddening histogram using the SFD98 reddening map. Two thirds of the QSOs are at low reddening of $E_{B-V}<0.03$.
\label{nz}}
\end{figure}

We furthermore restrict the sample to a magnitude range of $i<19.1$, where the follow-up of the SDSS candidate list is complete, and to a redshift range of $z=[0.5,2.5]$, where the SDSS colours are not affected by the spectrum short-ward of the Lyman break at 912\AA. We are thus left with 48,259 QSOs in our cleaned sample (see Fig.~\ref{nz} for a redshift histogram) and an additional 4,597 QSOs at Southern Galactic latitudes. Most QSOs in this sample are at very low reddening, as 65\% of them have $E_{B-V}<0.03$ while only 0.7\% have $E_{B-V}>0.1$.

According to the QSO target selection in the SDSS \citep{Rich02}, there is an exclusion zone for the stellar locus, in which QSO candidates would be outnumbered by stars and spectroscopic follow-up would become less efficient. This exclusion zone does affect our sample in the redshift range $z=[2.2,2.5]$ such that the red tail reaches the exclusion zone \citep[see Fig.~14 in][]{Rich02}, although they also claim that the sample is 90\% complete at $z<2.5$ (see their Fig.~10). A varying bias in the reddening could still lead to a variation in the exclusion of red QSOs. We find, however, that restricting the QSO sample to $z<2.2$ changed the results only randomly and within the size of our error bars.

\subsection{De-reddening stars and QSOs}

Starting from an extinction law we can derive bandpass-dependent absorption coefficients for specific SEDs. The non-zero width of a filter bandpass causes slightly different reddening for different SEDs, and the absorption coefficients cited in SF11 are derived for the SED of a star with $T_{\rm eff}=$7,000~K, $\log g=4.5$ and $[M/H]=-1$. As our study is entirely based on QSOs, we first determine the appropriate $A_{\rm band}$ values for QSOs using the SDSS bandpasses and the SDSS QSO template spectrum by \citet{vdB01}, using the extinction laws by \citet{F99} and \citet{OD94} as alternatives. While we use only one star spectrum at 7,000~K effective temperature to compare with literature, we calculate absorption coefficients for a whole grid in redshift and QSO continuum slope. We then average these values with a weighting corresponding to our QSO sample. 

As usual, we define bandpass-specific absorption coefficients $R_{\rm band} = A_{\rm band}/E_{B-V}$ and colour-specific reddening coefficients $R_{a-b} = R_a - R_b$. We find the expected absorption coefficients for 7,000~K stars and QSOs to be within $|\Delta R_{\rm band}|<0.01$ in the SDSS $griz$ bandpasses, but the $u$-band has a bluer effective wavelength when observing QSOs and thus an absorption coefficient larger by $\Delta R_u=+0.035$. This implies $R_{u-g,\rm QSO} \approx R_{u-g,F*}+0.035$, a correction we take into account when comparing to the literature.

\begin{figure}
\centering
\includegraphics[clip,angle=270,width=\hsize]{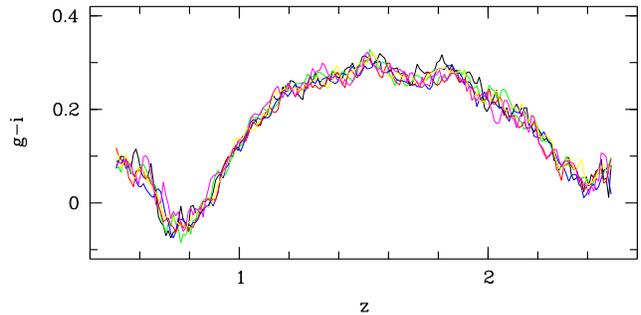}
\caption{QSO reference colour vs. redshift for six SDSS camera columns. The camera columns show variations in detector efficiency and filter curves.
\label{ridges}}
\end{figure}

\begin{figure}
\centering
\includegraphics[clip,angle=270,width=\hsize]{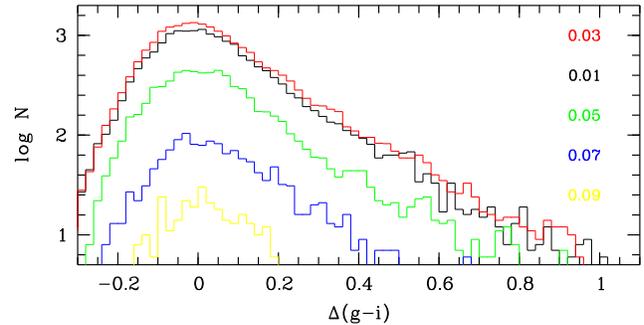}
\caption{Histogram of QSO colour residuals, i.e. dereddened colour minus reference colour, shown in reddening bins of full width $\Delta E_{B-V}=0.02$, centred on the values shown by the legend. The shape is best explained by a normal distribution that stems from the properties of the QSO accretion disk and a red tail that is caused by internal reddening in the host galaxy and its nucleus. Most relevant for this study is that the shape is intrinsic and has no variation correlated with foreground reddening, as expected.
\label{nresi}}
\end{figure}

\subsection{Redshift dependence of mean QSO colour}

The observed-frame colours of QSOs depend on the spectral slope of the QSO continuum light as well as on a contribution from emission lines that changes with redshift as the lines move through the filter curves. We thus determine empirically a QSO reference colour as a function of redshift, so that the following analysis can focus on relative colour offsets.

This reference colour is determined for every one of the six SDSS camera columns independently, as every camera column has its own set of detectors and filters with slightly varying transmission curves (see Fig.~\ref{ridges}). We determine the intrinsic colour of every QSO by subtracting the estimated reddening from the observed colour, using either the SFD98 or the SF11 reddening vector. The resulting QSO colour histogram at fixed redshift shows a steep edge on the blue side and a longer tail on the red side which is due to internal reddening by dust in the QSO nucleus or host galaxy (see Fig.~\ref{nresi}). For distributions with strong tails, the interquartile mean of a property (the sample mean obtained after excluding the two quartiles of most extreme objects) is a more robust estimate of the typical value than the regular mean considering the full distribution. But since the QSO colour distribution is tailed only on the red side, we choose to define our reference colour as the mean of the sample in a given redshift bin after excluding a quartile at the red end but only an octile at the blue end.

\begin{figure*}
\centering
\includegraphics[clip,angle=270,width=0.99\hsize]{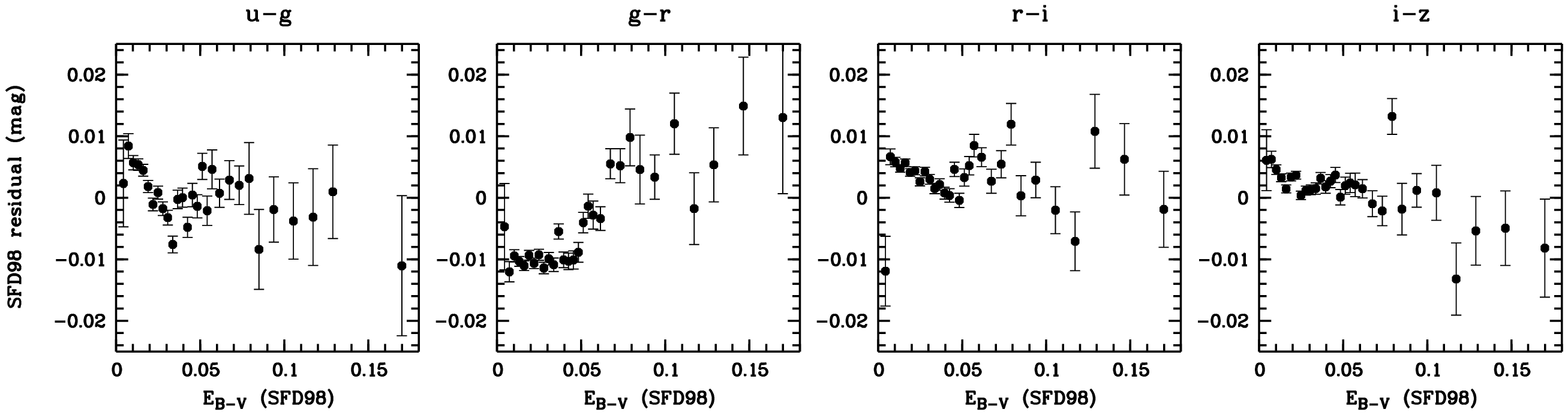}
\includegraphics[clip,angle=270,width=0.99\hsize]{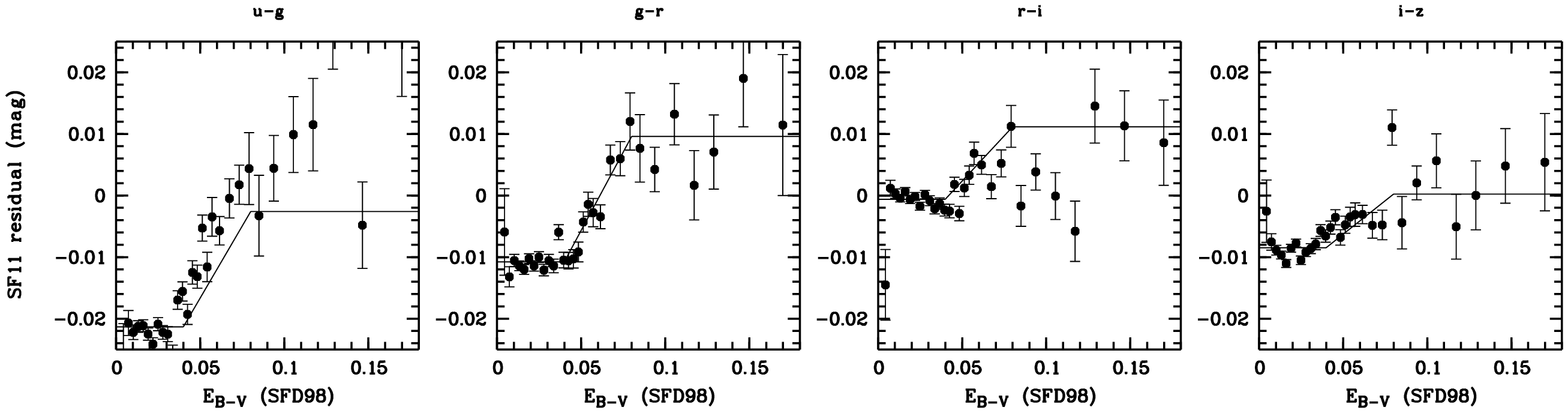}
\includegraphics[clip,angle=270,width=0.99\hsize]{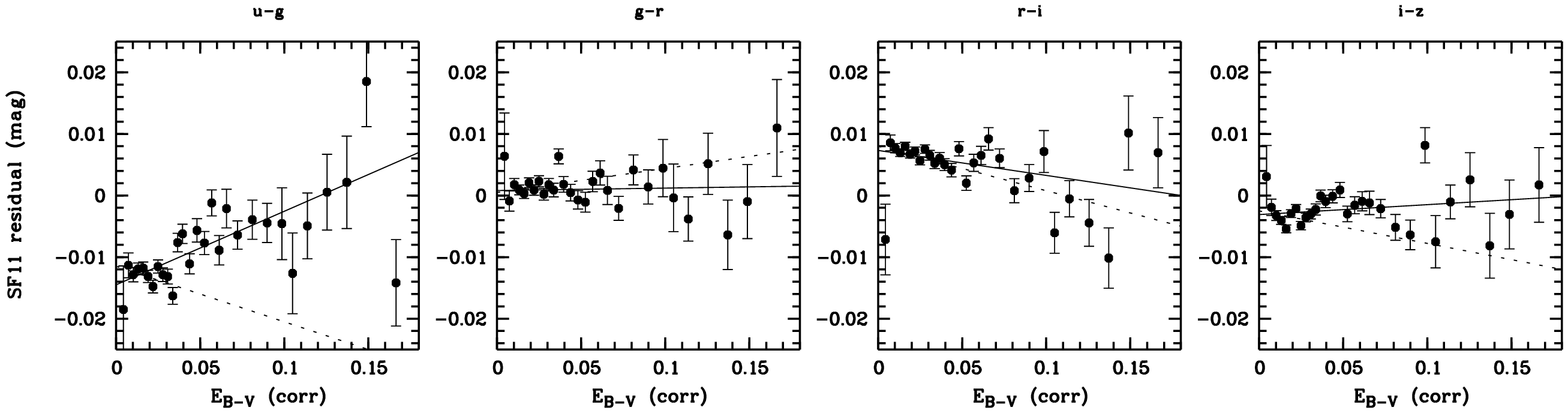}
\includegraphics[clip,angle=270,width=0.99\hsize]{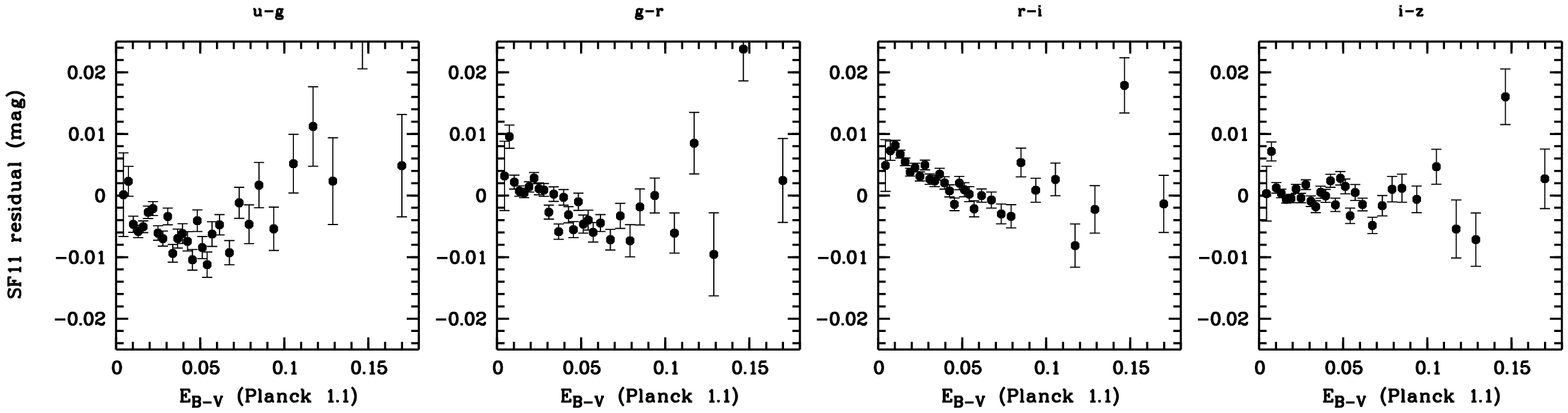}
\includegraphics[clip,angle=270,width=0.99\hsize]{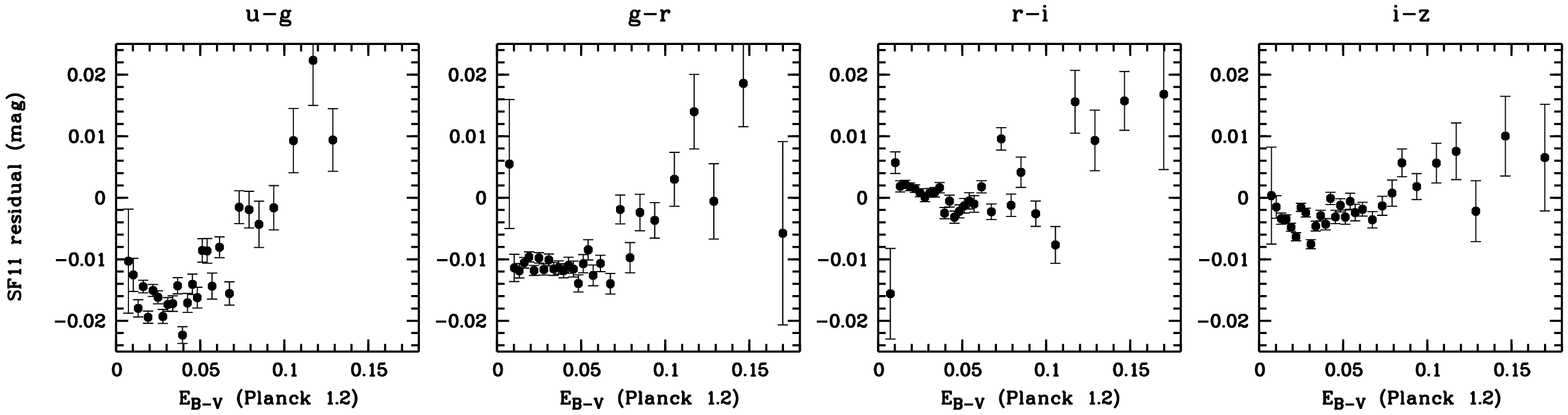}
\caption{Colour residuals of QSOs vs. reddening (arbitrary zeropoints), implying an imperfect reddening correction. {\it Top:} Using the SFD98 reddening map and vector.
{\it 2nd row:}  Using SF11 reddening vector with the SFD98 map (line is the predicted residual for the simple non-linearity correction). 
{\it 3rd row:}  SF11 vector and corrected reddening (see Tab.~1 for fit parameters).
{\it 4th row:}  SF11 vector and {\it Planck} V1.1 map.
{\it Bottom:}  SF11 vector and {\it Planck} V1.2 map.
\label{Colresi}}
\end{figure*}

\section{Colour residuals and the $E_{B-V}$ map}

We now determine residual colours for each QSO by subtracting the reference colour for the redshift and camera column of each QSO and deredden it with a reddening estimated from the $E_{B-V}$-map and reddening vector. We calculate the mean residual in bins of estimated reddening again using our variant of interquartile means that clips the reddest quartile and the bluest octile. If both the $E_{B-V}$-map and the reddening vector are correct, the residuals will show no trend with estimated reddening in any colour index. However, the following issues could produce residuals: 

\begin{enumerate}
\item A wrong reddening vector means that the true $R_{a-b}$ differs from the assumed one, and over- or undercorrecting the colour proportional to the reddening causes a trend of residual colour with a slope of $\Delta R_{a-b} = R_{a-b,\rm true}-R_{a-b,\rm assumed}$.

\item A wrong scale factor in the reddening map means that that the reddening vector appears to change in length but not direction, so that the $R_{a-b}$ values of all colour indices are stretched by the same factor including, consistently, $E_{B-V}$ itself.

\item Biasing the SFD map in a spatially variable manner means that lines-of-sight are moved to other reddening bins by variable amounts. This can manifest as a non-linear trend, where the scale factor of the SFD map appears to change with reddening before it stabilises at $E_{B-V}$-levels where the bias disappears in noise. The shape of the resulting non-linearity would be common to the residuals of all colours. A non-linearity appearing only in one colour index, however, would be best explained with a change in the reddening vector as we go from low-extinction regions in the sky to high extinction.
\end{enumerate}

Local biases of the reddening map could arise from a non-uniform extragalactic FIR background as suggested by \citet{Ya07} and \citet{Ka13}, from changes in the typical emissivity of dust that wasn't accounted for, i.e. a wrong temperature correction \citep[e.g.][]{PG10,S10} and from imperfect subtraction of zodiacal light. 

We note that in all previous investigations $R_{g-r}$ has always been close to 1, since reddening in $g-r$ and $B-V$ are intrinsically similar despite some difference in the wavelength coverage of these pairs of passbands. Hence, there is a limit to residual slopes in $g-r$ that are compatible with a correctly scaled $E_{B-V}$-map as long as the extinction law is smooth. 

In Fig.~\ref{Colresi} we show mean residuals in four SDSS colour indices as a function of estimated $E_{B-V}$ for different dust maps and reddening vectors.

\subsection{Using the SFD98 reddening map}

The top row of Fig.~\ref{Colresi} shows residuals after classic dereddening with SFD98, using their $E_{B-V}$-map and reddening vector. As expected, we see no slope in $g-r$, at least at $E_{B-V}<0.04$, but we see trends in the three other colour indices, suggesting the need to modify the reddening vector. Indeed, \citet{SF11} have already suggested an update, which we test in the next section.

\subsection{Using the SF11 reddening vector}

In the second row of Fig.~\ref{Colresi} we plot the colour residuals after dereddening with the SF11 reddening vector. We clearly see a non-linearity at $E_{B-V} \approx 0.06$, most evidently in $g-r$, but indeed in all colour indices, suggesting a need to modify the reddening map as well. The non-linearity in the $(g-r)$-colours of QSOs is very similar in shape and strength to that seen in galaxies by \citet{Peek13}, which we take as additional independent evidence of its nature as an issue with the reddening map. However, if this non-linear feature was restricted to the $(g-r)$-colour of QSOs, then it could not be seen as evidence for an issue with the dust map despite its coincidence with the \citet{Peek13} finding. Indeed, if it was absent from other colours it would more likely be an issue with the SDSS calibration that happened by chance to drift on the sky with the distribution of reddening. A change in extinction law with position in the sky could be hypothesised as well, although the effect looks too strong to be compatible with a correct $E_{B-V}$. The suggestion of a non-linearity common to all four investigated colours, however, supports very strongly the idea that it is indeed a shortcoming of the $E_{B-V}$-map.

We now describe the non-linearity in reddening to first order with a simple approach: we choose to (1) keep reddening unchanged in the interval $E_{B-V}=[0,0.04]$, (2) increase the strength of reddening by 50\% by remapping the interval $E_{B-V}=[0.04,0.08]$ onto $E_{B-V,\rm corr}=[0.04,0.10]$, and (3) keep the strength of additional reddening unchanged at $E_{B-V}>0.08$ by adding a constant $\Delta E_{B-V}=+0.02$ (see lines in Fig.~\ref{Colresi}). This description has no physical background and is only an ad-hoc fix. However, it appears to describe the trends in $g-r$ and $i-z$ with the correct strength. The match appears less good in $u-g$ and $r-i$, although this can also indicate a residual slope arising from a suboptimal reddening vector.

\subsection{Using a corrected map and the SF11 reddening vector}

In the third row of Fig.~\ref{Colresi} we assume the SF11 reddening vector again but apply the non-linear correction to the $E_{B-V}$-map. As a result we now get near-perfect residuals in $g-r$ and no significant overall non-linearities. However, despite the S11 reddening vector there remains a slope of $\Delta R_{r-i}=-0.04$, while in $u-g$ there is a suggestion of a deviation above $E_{B-V}=0.03$. We fit residual slopes and determine improved $R_{a-b}$ values for the low-reddening regime of $E_{B-V}<0.03$ and the overall range separately. Using simple least-squares fits to the mean residuals in the figure, we plot the two fits and list the results in Tab.~\ref{Rvalues} alongside the values of SFD98 and SF11, which are calibrated for 7,000~K stars (for a fair comparison we also list the $R_{u-g}$-value adjusted for 7,000~K stars). In summary we note that the SF11 vector appears to undercorrect reddening in $u-g$, although our reddening coefficient of $R_{u-g}=1.017$ is much closer to the SF11 value of $0.936$ than to the SFD98 value of $1.362$. Indeed, our value agrees perfectly with that measured from blue-tip stars by S10.

\subsection{Using the {\it Planck} maps}

The {\it Planck} Collaboration published $E_{B-V}$-maps \citep{Planck} in two versions, the first of which, V1.1, was compared to the SFD98 map by \citet{Peek13}: using the $g-r$ colours of galaxies they showed that the {\it Planck} map contained no such non-linearity as the SFD98 map does at $E_{B-V}<0.1$. We thus plot the full set of colour residuals of QSOs over the reddening from both versions V1.1 and V1.2 of the {\it Planck} map in the two bottom rows of Fig.~\ref{Colresi}. We use the SF11 reddening vector as it is close to our best-fitting $R_{a-b}$ values. We find indeed no sign of a non-linearity in the range of $E_{B-V}=[0.04,0.08]$, but instead observe a change in behaviour only above $E_{B-V}=0.07$, where the QSO data are noisier.

Furthermore, we observe significant slopes in all colours, including $g-r$, which suggest that the V1.1 map is scaled too high and leads to a general overcorrection of reddening in the very low-extinction regime. This shortcoming was evidently put right by the following version V1.2, where the reddening scale has been calibrated to fit the colours of QSOs and leave no residual by design in $g-r$ after dereddening. However, V1.2 also appears to contain a significant non-linearity again at $E_{B-V}>0.07$.

\subsection{Non-linearities in reddening maps compared}

\begin{figure*}
\centering
\includegraphics[clip,angle=270,width=\hsize]{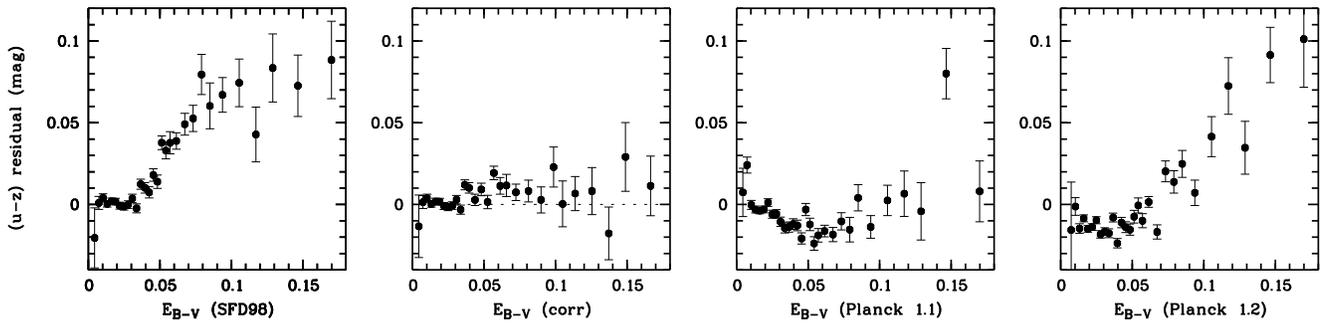}
\caption{Colour residuals in $u-z$ vs. reddening using four different $E_{B-V}$-maps and the SF11 dereddening vector. The original SFD98 map and the {\it Planck} V1.2 map appear to have a non-linearity changing the slope around $E_{B-V}\sim 0.05$. 
\label{umzresi}}
\end{figure*}

We finish this section by comparing the $E_{B-V}$-maps again with $u-z$ colour residuals. The colour index $u-z$ spans the full range of the SDSS filter set giving the maximum sensitivity to total absorption within the SDSS dataset \footnote{We repeated the exercise with $g-z$ to avoid possible calibration issues with the SDSS u-band, but found similar results.}. Also, the reddening coefficient we find, $R_{u-z}=3.04$, is remarkably close to a canonically assumed  $R_V=3.1$\footnote{
\citet{Berry12} and \citet{Mortsell13} constrain $R_V\approx 3$ from stars and quasars, resp., and find no evidence of large variations at higher-latitude $b$},
in which case we expect $A_V\approx R_{u-z}$. In Fig.~\ref{umzresi} we now deredden the data always with the SF11 vector, so that the only difference between panels is the choice of $E_{B-V}$-map. We clearly see the non-linearity of the classic SFD98 map (left panel), which is nearly removed by our simple approximation (2nd panel). The {\it Planck} map V1.1 shows only a weak non-linearity while the V1.2 map shows a non-linearity similar in strength to the SFD98 map, but appearing at slightly higher $E_{B-V}$.

\section{The SDSS North-South asymmetry}

S10 have reported a remarkable colour difference between blue-tip stars in the SDSS footprints at Northern vs. Southern Galactic latitudes. The degree of this difference is below $0\fm03$ for any colour index formed from a neighbouring pair in wavelength, but adds up to $0\fm049$ across the full range from $u$ to $z$, which would be slightly above the target calibration accuracy of better than $0\fm03$ in any band. A subsequent study (SF11) based on a wide variety of stars and intrinsic colours predicted by spectral properties found a milder calibration gradient with a $u-z$ colour difference of only $0\fm024$ between North and South. Possible origins of this difference in observed colour included differences in calibration, in dust properties and in the stellar population of the two Galactic hemispheres.

Here, we have probed the colour differences of QSOs between the two hemispheres and confirm a $u-z$ colour difference at the level of $\approx 0\fm03$ (see Tab.~\ref{NSasym}). However, the result is not unique given the uncertainty about the $E_{B-V}$ scale, which we subjected to an ad-hoc non-linearity correction. While $>80$\% of the Northern QSOs are at $E_{B-V}<0.04$, where we have not applied any correction, only a third of the already $10\times$ smaller Southern QSO sample is. We thus determine the colour difference between North and South for both a very-low reddening sample and a larger sample with the altered $E_{B-V}$ scale. The results are consistent within the errors, and are also broadly consistent with those of S10 and SF11, with the one exception that both we and S10 find the largest contribution coming from an offset in $g-r$, while SF11 find a smaller $g-r$ offset.

We consider it extremely unlikely that there is any hemispheric difference in $z>0.5$-QSO populations. Differences in dust properties are not ruled out, although we consider this option unlikely given the low levels of reddening and the fact that the largest colour difference is seen specifically in the $g-r$ colour, both for QSOs ($0\fm017$) and blue-tip stars ($0\fm022$). We thus propose that the predominant cause of the North-South asymmetry is a difference in the SDSS flux calibration, which could perhaps be confirmed by a comparison with Pan-STARRS photometry \citep{S12}.

\begin{table}
\caption{$R_{a-b}$ values derived from linear fits to residual colour trends: a significant change occurs in $R_{u-g}$ around $E_{B-V}\approx 0.04$. $R_{u-g}$ is also affected by the different source SEDs of QSOs vs. 7,000~K (F-type) stars as used by S10 and SF11 (their Table 6).
\label{Rvalues}}
\begin{tabular}{lcccc}
\hline \noalign{\smallskip}  
colour	&  $R_{a-b}$ 	& $R_{a-b}$ 	& $R_{a-b}$  		&  $R_{a-b}$  \\
		& SFD98		& SF11		& $E_{B-V}<0.03$ 	& all \\
\noalign{\smallskip} \hline \noalign{\smallskip}
u-g (qso) &  			& 			& $0.879 \pm 0.06$ & $1.052 \pm 0.02$ \\
u-g (F*)	& $1.362$ & $0.936 $ & $0.844 \pm 0.06$ & $1.017 \pm 0.02$ \\
g-r		& $1.042$ & $1.018 $ & $1.067 \pm 0.06$ & $1.027 \pm 0.02$ \\
r-i		& $0.665$ & $0.587 $ & $0.515 \pm 0.04$ & $0.547 \pm 0.02$ \\
i-z		& $0.607$ & $0.435 $ & $0.388 \pm 0.07$ & $0.451 \pm 0.02$ \\
\noalign{\smallskip} \hline
\end{tabular}
\end{table}

\section{Origin of non-linearities in the SFD98 map}

The SFD98 map has been tested for reliability using residual colours of stars (S10), galaxies \citep{Ya07,PG10,Peek13} and now also QSOs (this work). The most recent work on stars is based on Pan-STARRS1 \citep{S14}. Where they plot residuals vs. reddening levels, these works appear to agree in finding consistent features: 

\begin{itemize}
\item Fig.~5 of \citet{Ya07} as well as \citet{Peek13} show apparent $g-r$-residuals of galaxies over SFD reddening, which mimic in structure and magnitude the feature we find among QSOs before our simple non-linearity correction. 
\item Although the focus of the work using stars is on the higher-reddening regime, both S10 and \citet{S14} find a non-linearity also at the low levels of reddening discussed in this work. S10 plot similar results within the errors, but do not investigate them in detail; the \citet{S14} results are much more sensitive than S10 and show a non-linear difference between their $E_{B-V}$-map inferred from star colours and SFD98, which is very similar in magnitude and structure to our non-linearity correction, albeit smoother than our simplistic step-wise linear model.
\end{itemize}

\citet{S14} express surprise about this difference between their extinction map and SFD98 and say its cause is not understood; they find indeed nearly identical residuals whether they compare against the SFD98 map or the $\tau_{353}$-estimate of the Planck V1.2 map, which is an estimate of the dust optical depth. However, \citet{Ka13} have demonstrated that the $E_{B-V}$-map of SFD98 is locally and systematically elevated at the locations of galaxies, most likely due to the far-infrared emission of these galaxies and their clustered environment. By stacking the SFD map at the locations of galaxies detected in SDSS they could quantify this effect even for faint galaxies of $r\approx 20$. They also found that the added FIR flux did not originate solely from the galaxies, on which the stacking was centred, but had a significant contribution from the group environment including galaxies that were too faint to be seen in SDSS, but still expected to cluster with the primary galaxy. 

The effect moves the lines-of-sight near over-densities of galaxies to apparently larger SFD98-reddening, and \citet{Ka13} claim that it is strong enough to produce the apparent non-linearities in colour residuals with reddening seen by \citet{Ya07}. If the clustered FIR background from galaxies can thus explain the residuals of galaxies described by \citet{Ya07} and \citet{Peek13}, it can also explain similar residuals on the lines-of-sight to stars and QSOs. When \citet{Ka13} stacked the SFD98 images of QSOs, they found only a very weak effect that amounts to a bias of less than 1~mmag in $E_{B-V}$, and one might thus expect that QSOs should not display such non-linearities. However, we need to differentiate the contribution of an object to a bias in the map from the bias an object is subjected to because of other objects. In fact, there are several galaxies adding to the FIR background per resolution element in the SFD98 map (smoothing length $\sim 6\arcmin$). In the study by \citet{Ka13}, 12.6 million galaxies with $r<20.5$ over a 7,270 deg$^2$ footprint translates into an average of $\sim 14$ galaxies per SFD98 smoothing circle, and thus all possible lines-of-sight in the sky are subject to clustered random variations in the extragalactic FIR background.

While \citet{FDS00} discovered a mean extragalactic FIR background, it still varies on scales ranging from much less than their map's resolution to larger than that. The FIR emission of distant galaxies is localised mostly by the extend of these galaxies, which is much smaller than the resolution of the SFD map, while the clustering of galaxies arranges their smoothed contribution into objects that are potentially larger than the resolution. Without prior knowledge of the locations of the galaxies SFD98 were thus not able to subtract this structured background. Any resolution element of the dust emission map thus contains a smooth contribution from Milky Way dust, which corresponds to an amount of extinction that represent the average over the lines-of-sight, plus a contribution from much smaller sources with a very small areal coverage that is being smoothed across the resolution element of the map and contaminates the extinction estimate of those lines-of-sight that look past the emitting small and distant galaxies.

It is therefore natural that any line-of-sight including those to stars and QSOs are fully affected by the biases in the reddening map that result from the unresolved but structured extragalactic background. The smoothing length of the SFD98 map makes the extinction bias along the lines-of-sight to galaxies non-specific and carries the bias to nearby lines-of-sight to stars and QSOs alike, and while QSOs may add only little to the bias, they are still fully affected by the bias introduced by galaxies. 

The features of the reddening map can thus be probed with any of these three tracers, which have different (dis-)advantages: stars are common and their photometry is straightforward, but they reside at different distances and may only be subject to part of the integrated line-of-sight extinction; galaxies are common, but may have clustering-dependent colours and their photometry can be biased by internal colour gradients and variation in seeing; QSOs may be the least biased probe, but they are rare and require averaging over wider areas.

\section{Conclusion}

We have used QSOs at redshifts $0.5<z<2.5$ from the SDSS DR7 QSO Catalogue \citep{DR7QSO} to investigate reddening due to Milky Way dust at low extinction of $E_{B-V}<0.15$. After dereddening and correcting for redshift with a function specific for each SDSS camera column, we find the dispersion of QSO $u-z$ colours to be $0\fm14$, independent of redshift and reddening. This statistic is obtained after slicing off the bluest octile and the reddest quartile to remove outliers and suppress noise from internal reddening of QSOs. We can thus determine mean colour residuals in any subsample with errors of $0\fm14/\sqrt{N}$ achieving a precision of 2 to 30 mmag in $u-z$ for our chosen reddening bins.

We confirm that the SF11 reddening vector is a big improvement over the SFD98 vector, although we prefer a slightly higher $u$-band extinction with a reddening coefficient $R_{u-g}=1.017$ instead of $0.94$, in perfect agreement with that measured from blue-tip stars by S10. We also confirm an apparent non-linearity in the $E_{B-V}$-scale of the SFD98 reddening map, which is very similar to that found by \citet{Ya07} and \citet{Peek13} using the colour residuals of galaxies, and which is also consistent with differences between the SFD98 map and the Pan-STARRS1 reddening map that was derived from star colours. 

A possible explanation for this behaviour was given by \citet{Ka13}, who show that the unresolved FIR background from galaxy groups introduces a faint but highly structured signal into the SFD98 map, which, after smoothing, affects all lines-of-sight, whether they point to the galaxies making the bias or instead to stars and QSOs on nearby lines-of-sight.

Assuming the SF11 extinction coefficients, residuals in $u-z$ are a very good approximation of $A_V$-residuals. Using these residual measures we compare the linearity properties of several reddening maps including those by the {\it Planck} team. We find that the {\it Planck} V1.1 map has no strong non-linearity in $E_{B-V}$, but seemingly a wrong $E_{B-V}$-scale, while V1.2 has by design the right scale below $E_{B-V}<0.07$ (as it is calibrated with QSO colour residuals in this regime), but again a non-linearity above. 

We conclude that the colours of extragalactic objects are useful constraints on the photometric calibration and total Milky Way reddening corrections in wide-area surveys. They can be used to constrain the mean precision of colours over wide areas to milli-magnitude levels and are useful to zoom into the properties of reddening in the low-extinction regime at $E_{B-V}<0.03$. We also showed that the North-South asymmetry seen in SDSS blue-tip stars by S10 is replicated in the colours of QSOs. In the future, we intend to use colour residuals of QSOs also in the ongoing SkyMapper Southern Survey to verify calibration and reddening.

\begin{table}
\caption{Colour difference South-minus-North measured after dereddening with SF11 law, using only QSOs at $E_{B-V}<0.04$ or including QSOs up to $E_{B-V}<0.1$ on the scale corrected for non-linearity; our values are broadly consistent with two previous results from SDSS stars (S10, SF11). 
\label{NSasym}}
\begin{tabular}{lrrrr}
\hline \noalign{\smallskip}  
colour	&  $\Delta c$ (mmag) &  $\Delta c$ (mmag) & Blue tip & Spectrum \\
		& $E_{B-V}<0.04$	& $E_{B-V}<0.1$	& S10	& SF11 \\
\noalign{\smallskip} \hline \noalign{\smallskip}
u-g	& $+0.7  \pm 1.6$	& $+1.2   \pm 1.1$	& +7.6	& $+2.3\pm7.2$  \\
g-r	& $+15.3\pm 1.3$ 	& $+17.3 \pm 0.9$	& +21.8	& $+8.8\pm1.5$  \\
r-i	& $+4.0  \pm 1.0$ 	& $+4.0   \pm 0.7$	& +7.2	& $+3.4\pm1.9$  \\
i-z	& $+9.4  \pm 1.1$ 	& $+11.9 \pm 0.7$	& +12.4	& $+9.3\pm1.6$  \\
u-z	& $+30.3\pm 3.3$ 	& $+34.7 \pm 2.2$	& +49.0	& $+23.8\pm 7.7$  \\
\noalign{\smallskip} \hline
\end{tabular}
\end{table}

\section*{acknowledgements}
Valuable discussions with David Nataf led to the preparation of this paper.
Funding for the Sloan Digital Sky Survey (SDSS) and SDSS-II has been provided by the Alfred P. Sloan Foundation, the Participating Institutions, the National Science Foundation, the U.S. Department of Energy, the National Aeronautics and Space Administration, the Japanese Monbukagakusho, and the Max {\it Planck} Society, and the Higher Education Funding Council for England. The SDSS Web site is http://www.sdss.org/. The SDSS is managed by the Astrophysical Research Consortium (ARC) for the Participating Institutions. The Participating Institutions are the American Museum of Natural History, Astrophysical Institute Potsdam, University of Basel, University of Cambridge, Case Western Reserve University, The University of Chicago, Drexel University, Fermilab, the Institute for Advanced Study, the Japan Participation Group, The Johns Hopkins University, the Joint Institute for Nuclear Astrophysics, the Kavli Institute for Particle Astrophysics and Cosmology, the Korean Scientist Group, the Chinese Academy of Sciences (LAMOST), Los Alamos National Laboratory, the Max-Planck-Institute for Astronomy (MPIA), the Max-Planck-Institute for Astrophysics (MPA), New Mexico State University, Ohio State University, University of Pittsburgh, University of Portsmouth, Princeton University, the United States Naval Observatory, and the University of Washington.

\end{document}